\documentclass[8pt,prd,superscriptaddress,nofootinbib,notitlepage]{revtex4-1}

\usepackage{slashed}
\usepackage{caption}
\usepackage{subfigure}
\usepackage{amsmath,amssymb,graphicx,xcolor,mathtools}
\usepackage[colorlinks=true,linktocpage=green,linkcolor=blue,citecolor=blue]{hyperref}
\numberwithin{equation}{section}
\allowdisplaybreaks
\usepackage{hyperref}
\hypersetup{
	colorlinks=true,
	citecolor = blue,
	linkcolor= blue,
	filecolor= blue,      
	urlcolor= blue,
	pdftitle={HDL}
}
\usepackage{float}
\usepackage{graphics}
\usepackage{mathrsfs}
\usepackage{simpler-wick}
\usepackage{dsfont}
\usepackage{mathtools}
\usepackage{appendix}

\newcommand{\ba}{\begin{eqnarray}}
\newcommand{\ea}{\end{eqnarray}}

\usepackage{caption}

\begin{document}
	
	\title{Thermoelectric transport in
   non-extensive QCD matter in presence of magnetic field }
	
		\author{Lateef Ahmad Wani}
	\email{lwani@kfu.edu.sa}
	\affiliation{Department of Mathematics and Statistics, College of Science, King Faisal University, P.O. Box 400,
Al-Ahsa 31982, Saudi Arabia}

	\author{Salman Ahamad Khan}
	\email{salmankhan.dx786@gmail.com}
	\affiliation{Department of Physics, Integral University, Lucknow - 226026, India}

\begin{abstract}
We investigate the thermoelectric response of a magnetized non-extensive quark–gluon plasma within the framework of kinetic theory using the relaxation time approximation (RTA). The interactions among the partons are incorporated through a quasiparticle model, where the medium-dependent quark mass is obtained from the poles of the resummed hard-thermal-loop (HTL) propagator. Our results show that non-extensive effects suppress the Seebeck coefficient both in the absence and in the presence of an external magnetic field. In contrast, the Nernst coefficient, which arises only in the presence of a magnetic field, is found to be significantly enhanced by non-extensivity. These findings demonstrate that deviations from equilibrium can substantially modify the thermoelectric transport properties of the quark gluon plasma. 
\end{abstract}

\maketitle
\vskip 0.01in

\begin{flushright}
{\normalsize
}
\end{flushright}

\section{Introduction}
{ A strongly interacting matter having quarks and gluons as fundamental degrees of freedom is produced in the ultra-relativistic 
heavy ion collisions at  RHIC (BNL USA) and LHC (CERN). The study of this extreme phase of matter is of paramount importance among the heavy ion physics community due to its great phenomenological importance in particle and astrophysics.  One of the major themes of the current research in nuclear theory has been the transport properties of this many body interacting relativistic system because transport properties are crucial inputs in the hydrodynamical modeling of the evolving matter.  In non-central collisions, a magnetic field of the order of $m_\pi^2$
at RHIC
\cite{Kharzeev:NPA803'2008} and  that of  $15m_\pi^2$ at 
LHC \cite{Skokov:IJMPA24'2009}) is also 
produced which modifies the transport phenomena in hot QCD matter. Hence,  the transport properties of magnetized strongly interacting matter  have been extensively explored using lattice QCD, pertubative QCD and various effective models~\cite{Rath:PRD102'2020,Kurian:EPJC79'2019,Li:PRD97'2018,
Hattori:PRD96'2017,Chen:PRD101'2020,Nam:PRD87'2013,
PV:PRL105'2010,Seung:PRD86'2012,Kharzeev:PPNP75'2014,
Satow:PRD90'2014,Pu:PRD91'2015,Hattori:PRD94'2016,
Khan:PRD104'2021,Khan:PRD106'2022}. \par

{ Another transport coefficient that measures a material's capacity to transform a temperature gradient into an electric current is the Seebeck coefficient. Over the years, the materials' thermoelectric characteristics have mostly been examined in relation to condensed matter physics~\cite{Pao:9505002,Matusiak:PRB97'2018,Gaudart:PSS2'2008,Wysokinski:JAP113"2013, Wojcik:PRB89'2014, Seo:PRB90'2014}. In the strongly 
interacting matter produced 
in the heavy-ion collisions,   there is a temperature difference between the fireball's central and peripheral region. In addition to it,
a finite baryon chemical potential is also needed to 
observe the thermoelectric effect in strongly 
interacting matter unlike the condensed matter systems since both the positive as well as negative charge participate in the transport process while in condensed matter system only single type of charge takes part. The Seebeck effect in strongly interacting matter has been investigated both in  absence~\cite{Bhatt:PRD99'2019,Aman:EPJC82'2022} as well as in the presence of magnetic field~\cite{Das:PRD102'2020,Kurian:PRD103,Zhang:EPJC81'2021,
Dey:PRD102'2020,Dey:PRD104'2021,Khan:PRC110'2024,
Shaikh:PRD11'2025,Gabuzyan:2025ben,Rath:PRD113'2026}.
  \par
  
  {The QGP fireball created in the heavy ion collision has a finite volume and has some long range correlations. The QGP phase is also influenced by the memory effects.  In all the studies mentioned above, these effects have been neglected.  One of the approach to take these effects  into account is the use of non-extensive Tsallis statistics  which is a generalization of the Maxwell Boltzmann statistics~\cite{Tsallis,Tsallis1}. Systems  having
long range interactions and correlations does not follow the equilibrium thermal distribution rather 
nonextensive particle distributions. It is reported
that the distribution of particles in space plasmas and laboratory plasmas do 
not exactly follow the  Maxwellian distribution~\cite{Liu,Pierrard,maksimovic}. In non-extensive statistics, entropy becomes a non-additive quantity and a  parameter $q$ which measures  the degree of non-extensivity in the system, is introduced in the distribution function. The Tsallis distribution approaches to the equilibrium Fermi-Dirac distribution when the parameter 
$q$ approaches to unity. In the context of heavy ion collisions, the non-extensive statistics have been extensively used to study the transverse momentum spectra and radial flow  of the final state hadrons ~\cite{Abelev:PRC75:2007,Adare:PRD83'2011,Adare1:PRC83'2011,
Aamodt:PLB693'2010,Aamodt:EPJC71'2011,Khachatryan:JHEP02'2010,
Khachartryan:PRL105'2010,Aad:NJP13'2011,Abelev:PRL109'2012,
wilk:PRL84'2000,shao:JPG37'2010,Wong:PRD87'2013,
Che:JPG48'2021,Tang:PRC79'2009}. The hydrodynamic equations have been derived for non-extensive medium  in the kinetic theory framework~\cite{osada:prc77'2008,biro:EPJA48'2009,biro:PRC85'2012}. Some authors have studied the non-extensive effects on charge, heat and momentum transport coefficients  of QGP in absence as well as in the presence of background magnetic field~\cite{Rath:EPJC83'2023,Rath:EPJA60'2024}.
  Alqahtani et
al. have derived the fluid dynamical equation for the anisotropic systems considering the nonextensive distribution of the partons and found that non-extensive effects modify the  bulk pressure evolution~\cite{alqahtani:EPJC82'2022}. In addition to these works, 
the energy loss, jet quenching parameter  and the nuclear modification factor  in the QGP have been explored  using the
nonextensive statistics~\cite{Tripathy:EPJA52'2016,Bhattacharya:PhysA624'2023,Bhattacharyya:PLB856'2024}. Authors in~\cite{sarwar:EPJC82'2022} used the non-extensive equation of state to study the propagation of nonlinear waves in quark gluon plasma in the framework of second order dissipative hydrodynamics. In an another work, the thermodynamic variables have been computed analytically
for a non-extensive  hot and dense system~\cite{Bhattacharya:PRD94'2016}. The effect of the non-extensivity on the properties of the magnetized  QCD medium have been explored 
in the NJL model in~\cite{Islam:EPJA60'2024}. Similarly, authors in~\cite{Xiang:EPJC85'2025} studied the properties of the  QCD phase diagram in the context of SU(2) NJL model having a finite  chiral chemical potential. Recently, the longitudinal and transverse dielectric functions of the hot QCD medium have been computed to study the refractive index of the medium~\cite{Jiang:PRD111'2025}. \par
{In this work, we study the effect of non-extensivity on the thermoelectric transport in the hot and dense strongly interacting matter produced in the heavy ion collision experiments. For that purpose, we have exploited the Tsallis non-extensive statistics, in place of conventional Fermi-Dirac statistics adopted in previous studies. The quark gluon plasma produced in the relativistic heavy-ion collisions is expected to exhibit temperature fluctuations, long-range correlations, and deviations from equilibrium which makes the use of the non-extensive statistics more realistic.  Furthermore, the results are compared with those obtained using the equilibrium Fermi-Dirac distribution to quantify the effects of the non-extensivity. To the best of our knowledge, this is  the first comprehensive study of thermoelectric transport coefficients  in a non-extensive medium and in the magnetic field providing new insights into transport phenomena in systems that deviate from local equilibrium.}

The paper is organized as follows: In section~\ref{two1}, we have derived the expression for the Seebeck coefficient in the absence of the magnetic field. In section~\ref{two2}, we computed the Seebeck and Nernst coefficients in the presence of the weak magnetic field. In section~\ref{three1} and section~\ref{three2}, we have analyzed  the results in absence and in the presence of the magnetic field, respectively.  finally, we conclude in section~\ref{four}.

\section{Theromoelectric transport coefficients  of a thermal QCD medium}\label{two}
 In the kinetic theory approach, the evolution 
of the phase space 
distribution function is
given  by relativistic Boltzmann transport equation (RBTE), which reads as
\begin{eqnarray}
p^{\mu}\frac{\partial f}{\partial x^{\mu}}+q~F^{\rho \sigma}
p_{\sigma}\frac{\partial f}{\partial p^{\rho}}= 
-\frac{p^\mu u_\mu}{\tau} \left(f-f_{0} 
\right),
\label{rbte}
\end{eqnarray}
where $f=f_{0}+\delta f$; $\delta f$ is  
small deviation from the equilibrium
distribution function $f_0$ and $F^{\rho \sigma}$
 corresponds to the electromagnetic field 
strength tensor. $\tau$ is the relaxation time of the particle.

\subsection{Seebeck coefficient in the 
absence of the magnetic field}\label{two1}
In this subsection, we evaluate the Seebeck coefficient of a thermal QCD medium consisting of up ($u$), down ($d$), and strange ($s$) quarks together with their corresponding antiquarks. Thermoelectric phenomena arise from the coupling between charge and heat transport in the presence of a temperature gradient. When a spatial temperature gradient is established across the medium, the charge carriers tend to diffuse from the hotter region toward the colder region due to their enhanced thermal motion. Consequently, an electric current is induced in the system. Under open-circuit conditions, however, the accumulation of charges generates an electric field that opposes the diffusion current. The balance between these two competing mechanisms gives rise to a measurable voltage difference, known as the Seebeck effect, and the corresponding proportionality constant is defined as the Seebeck coefficient.

The induced electric current density in the presence of an external electric field and a temperature gradient can be written as
\ba
J_{\mu}=\sum_{i} g_i \int 
\frac{d^3p}{(2\pi)^3}~\frac{p_{\mu}}{\omega_i} 
~(q_i\delta f_i+\bar{q_i}\delta \bar{f}_i),
\label{current_temp}
\ea
where $\delta f_i$ and $\delta \bar{f}_i$ are 
 the infinitesimal deviations  in the quark and anti quark distribution functions, respectively and $g_i$ refers to 
the degeneracy factor.\par
The Boltzmann transport equation~\eqref{rbte} 
in the presence of electric field and 
temperature gradient reads as 
\ba
\vec{p}.\frac{\partial f}{\partial {\vec{r}}}+q ~
{\vec {E}.\vec{p}}
 \frac{\partial f}{\partial p^0}+
q p_0 ~{\vec{E}.}\frac{\partial f}{\partial {\vec{ p}}}=
-\frac{p^{\mu}u_{\mu}}{\tau} \left(f-f_{\mathrm T}
\right),
\label{rbte_temp}
\ea
where $f=f_{\mathrm T}
+\delta f$. $\delta f$ is the deviation in the Tsallis distribution function since the nonextensive systems can also get further
deviated from equilibrium due to the presence of external forces or fields. In the absence of any external perturbation, the distribution of quarks (anti-quarks) in the medium is governed by the Tsallis distribution function
which is given by~\cite{cleymans:JPG39'2012,conroy:PRD78'2008,
Biro:PRC85'2012}
\ba 
f_{\mathrm T}(p,q)=\frac{1}{[1+(q-1)\beta(\omega \mp\mu)]^{\frac{1}{(q-1)}}+1} ,
\ea
where $-(+)$ signs corresponds to the quarks (anti-quarks). $q$ is the non-extensive parameter which measures  the departure of the system from the equilibrium. Tsallis distribution function is generalization of the Fermi-Dirac distribution function. As the nonextensive parameter $q$ approaches unity, the Tsallis distribution  reduces to the Fermi-Dirac distribution 
\ba 
f(p)=\frac{1}{e^{\beta(\omega \mp\mu)}+1}.
\ea

The RBTE~\eqref{rbte_temp}
can be recast after some simplification as 
\ba \label{appendix_A}
\delta f&=&
\frac{\vec{p}}{\omega_i}.(\omega-\mu)\tau \left(-\frac{1}{T^2}\right)~\{1+(q-1)\beta(\omega-\mu)\}^{-1}
f\left( 
1-f\right)~\nabla_{\vec{r}}
 T(\vec{r})~ \nonumber\\
&&+2q\beta \tau~ \frac{{\vec{E} \cdot \vec{p}}}
{\omega} ~\{1+(q-1)\beta(\omega-\mu)\}^{-1}f_{\mathrm T}\left( 
1-f_{\mathrm T} \right).  
\ea
Similarly, we compute the deviation in the anti quark distribution function  $\delta \bar{f}$ as 
\begin{eqnarray}
\delta \bar{f}&=&\frac{\vec{p}}{\omega}.(\omega +\mu)
\tau \left(-\frac{1}{T^2}\right)~\{1+(q-1)\beta(\omega +\mu)\}^{-1}
\bar{f_{\mathrm T}}\left( 
1-\bar{f_{\mathrm T}} \right)\nabla_{\vec{r}} T(\vec{r}) \nonumber \\
&& +2\beta \bar{q} \tau
\frac{\vec{E}.\vec{p}}{\omega}\{1+(q-1)\beta(\omega +\mu)\}^{-1}  ~\bar{f_{\mathrm T}}(1-\bar{f_{\mathrm T}}).
\label{eqoneseven}
\end{eqnarray}
With the help of  $\delta f$,
$\delta \bar{f}$ and Eq.~\eqref{current_temp}, we obtain the induced current
due to a single quark flavor  which is given by
\ba
J_{k} &=&   q g\tau\int 
\frac{d^3p}{(2\pi)^3}~ 
\bigg [\bigg \{\frac{p_k^2}{\omega^2}(\omega-\mu) \left(\frac{-1}{T^2}\right)~\{1+(q-1)\beta(\omega-\mu)\}^{-1}
f_{\mathrm T}\left( 
1-f_{\mathrm T} \right)\nabla_{\vec{r}} T(\vec{r}) \nonumber\\
&& +2\frac{p_k^2}{\omega^2}
 qE_k\beta ~\{1+(q-1)\beta(\omega-\mu)\}^{-1}  f_{\mathrm T}
(1-f_{\mathrm T})\bigg \} \nonumber\\
&& + \bar{q} g \tau \int 
\frac{d^3p}{(2\pi)^3}~ 
\bigg[\bigg \{\frac{p_k^2}{\omega^2}(\omega +\mu) \left(\frac{-1}{T^2}\right)~\{1+(q-1)\beta(\omega +\mu)\}^{-1}
\bar{f_{\mathrm T}}\left( 
1-\bar{f_{\mathrm T}} \right)\nabla_{\vec{r}} T(\vec{r}) \nonumber \\
&& +2\frac{p_k^2}{\omega^2}
 \bar{q}E_k\beta ~\{1+(q-1)\beta(\omega +\mu)\}^{-1} \bar{f_{\mathrm T}}
(1-\bar{f_{\mathrm T}})\bigg \}.
\label{induced_c}
\ea

{ Under the open-circuit condition, the net electric current associated with the $i^{\rm th}$ quark flavor vanishes, i.e.,
$\mathbf{J}_i=0$
In this situation, the thermally induced diffusion current is exactly balanced by the drift current generated by the induced electric field. Consequently, no net charge transport occurs in the medium, and a steady-state equilibrium is established. By setting the induced current  equal to zero, one obtains the  relation between the electric field and the temperature gradient
as
\begin{equation}
\mathbf{E}=S_i\,\nabla T,
\end{equation}
where $S_i$ corresponds to the Seebeck coefficient corresponding to the $i^{\rm th}$ quark flavor. This relation serves as the foundation  for evaluating the thermoelectric transport coefficient of each quark flavour within the kinetic theory framework. Therefore, we put the induced 
current~\eqref{induced_c} to zero such that 
\begin{eqnarray}
{\vec{E}} &=&\frac{1}{2 Tq}\left(\frac{M_1+M_2}{M_3+M_4}\right) 
~{ \nabla_{\vec{r}}} T(\vec{r}),\nonumber \\
&\equiv& S~{\nabla_{\vec{r}}} T(\vec{r}).
\label{sstate_b0}
\end{eqnarray}
Here $S$ is the Seebeck coefficient, which
is given by}
\ba
S= \frac{1}{2 Tq}\left(\frac{M_1+M_2}{M_3+M_4}\right),
\ea 
where 
\begin{eqnarray}
M_1&=&\frac{1}{3}\int \frac{d^3p}{(2\pi)^3}~ 
\frac{p^2}{\omega^2}(\omega -\mu)~\{1+(q-1)\beta(\omega-\mu)\}^{-1} 
f_{\mathrm T}\left( 
1-f_{\mathrm T} \right),\\
M_2&=&-\frac{1}{3}\int \frac{d^3p}{(2\pi)^3}~ 
\frac{p^2}{\omega^2}(\omega +\mu)~
\{1+(q-1)\beta(\omega +\mu)\}^{-1} 
\bar{f_{\mathrm T}}\left( 
1-\bar{f_{\mathrm T}} \right),\\
M_3&=&\frac{1}{3}\int \frac{d^3p}{(2\pi)^3}~ 
\frac{p^2}{\omega^2}~\{1+(q-1)\beta(\omega-\mu)\}^{-1} 
f_{\mathrm T}\left( 
1-f_{\mathrm T} \right),\\
M_4&=&\frac{1}{3}\int \frac{d^3p}{(2\pi)^3}~ 
\frac{p^2}{\omega^2} ~\{1+(q-1)\beta(\omega +\mu)\}^{-1}
\bar{f_{\mathrm T}}\left( 
1-\bar{f_{\mathrm T}} \right).
\end{eqnarray}
Since we have only looked at one quark flavour thus far, we will now concentrate on the hot QCD medium with several quark flavours. In this instance, we have taken into consideration quark gluon plasma with three flavours (u, d, and s quarks and their anti-particles). Due to individual flavours, the overall induced current can be expressed as the sum of the currents.
\ba
\vec{J}&=& \sum_{i}\vec{J}_i \nonumber\\
&=&\left(\frac{q^2_1g_1\tau_1}{T}(M_3+M_4)_1+
\frac{q^2_2g_2\tau_2}{T}(M_3+M_4)_2+....\right)\vec{E}\nonumber\\
&&-\left(\frac{q_1g_1\tau_1}{T^2}(M_1+M_2)_1+
\frac{q_2g_2\tau_2}{T^2}(M_1+M_2)_2+.....\right)
\nabla_{\vec{r}} T(\vec{r}).
\ea
In the steady state condition, the total induced 
current vanishes {\em i.e.} $\vec{J}=0$. As a result, we get
\ba
\vec{E}= \frac{1}{2T}\frac{\sum_i q_ig_i\tau_i(M_1+M_2)_i}
{\sum_i q^2_ig_i\tau_i (M_3+M_4)_i}\nabla_{\vec{r}} 
T(\vec{r}),
\label{tot_current}
\ea
which gives the seebeck coefficient for the 
medium as
\ba
S_{\rm tot}=\frac{1}{2T}\frac{\sum_{i} q_ig_i\tau_i(M_1+M_2)_i}
{\sum_i q^2_ig_i\tau_i (M_3+M_4)_i}.  
\ea
Since each flavour has the same degeneracy factor and relaxation time, the medium's overall Seebeck coefficient can be written in terms of each flavor's Seebeck coefficient as
\ba
S_{\rm tot}=\frac{\sum_i S_i q^2_i(M_3+M_4)_i}
{\sum_i q^2_i (M_3+M_4)_i}.  
\ea
In the limit when non-extensive parameter $q$ approaches to unity, the above derived expression for the Seebeck coefficient
takes the form obtained in~\cite{Dey:PRD102'2020} for the extensive QCD medium where the distribution of particles in the medium is governed by the Fermi-Dirac statistics.

\subsection{Seebeck and Nernst coefficients in the
presence of the magnetic field}\label{two2}

{ In the present work, the weak magnetic field approximation is considered, where the magnetic field strength satisfies the scale hierarchy
$
eB \ll T^{2}.
$ Under this condition, the spacing between the Landau levels is much smaller than the characteristic thermal energy of the particles, resulting in the occupation of a large number of Landau levels. Consequently, the discrete Landau spectrum effectively approaches a quasi-continuous momentum distribution, and the effects of Landau quantization become negligible. Therefore, the phase-space integration is performed using the conventional three-dimensional momentum measure,
instead of replacing it by a summation over Landau levels. This approximation is well justified within the framework of relativistic kinetic theory and has been extensively used in the estimation of transport properties of weakly magnetized quark gluon plasma~\cite{Dey:PRD104'2021,Khan:PRC110'2024,Rath:EPJC83'2023,
Feng:PRD96'2017}. In this regime, the magnetic field influences the transport coefficients perturbatively via the Lorentz force, while the quantization of the transverse motion can be safely neglected.}
 In the weak magnetic field, the dispersion relation of the 
charged particle is not directly affected by the 
 magnetic field rather $B$ acts as a perturbation.
 The induced four current in the medium is given by
\begin{eqnarray}
J_{\mu}=\sum_{i}g_i \int \frac{d^3p}
{(2\pi)^3}\frac{p_{\mu}}{\omega_i}
[q_i\delta f_i+\bar{q}_i\delta \bar{f}_i],
\label{current_weak}
\end{eqnarray}
where $\omega_i=\sqrt{p^2+m_i^2}$. 
The RBTE~\eqref{rbte} in the presence of the 
Lorentz force can be written as (see the Appendix A)
\begin{eqnarray}
\frac{\partial f}{\partial t}+\vec{v}.\frac{\partial f}{\partial \vec{r}}+
\vec{F}. \frac{\partial f}{\partial \vec{p}}&=&
-\frac{1}{\tau}\left(f-f_{T} \right),
\label{rbte_lorenz}
\end{eqnarray}
where $\vec{F}=q~(\vec{E}+\vec{v}\times \vec{B})$ is the Lorentz force. In this study, we have assumed that the direction of the the electric field is along the  $xy$ plane {\em i.e.}
$\vec{E}=E_x \hat{x}+E_y \hat{y}$
and that of magnetic field is along $z$ direction {i.e} $\vec{B}=B\hat{z}$. Then
for QCD medium, which is homogeneous in time, Eq.~\eqref{rbte_lorenz} can be written as 
\begin{eqnarray}
q_fB\tau \left(v_x\frac{\partial f}{\partial p_y}-
v_y \frac{\partial f}{\partial p_x}\right)-\tau
\vec{v} .\frac{\partial f}{\partial \vec{r}}-
\tau q_f \vec{E} .\frac{\partial f}{\partial \vec{p}}=
\delta f,
\label{43}
\end{eqnarray}
In order to solve Eq.~\eqref{rbte_lorenz}, 
we take an ansatz~\cite{Feng:PRD96'2017}
\begin{eqnarray}
\delta f= f-f_{T}= -\tau q\vec{E} 
.\frac{\partial f_{T}}{\partial \vec{p}}
-\vec{\lambda}. \frac{\partial f_{T}}{\partial \vec{p}}
\label{ansatz}
\end{eqnarray}
Now, equating  Eqs.~\eqref{43} and \eqref{ansatz},
we get
\begin{eqnarray}
\vec{\lambda}. \frac{\partial f_{T}}{\partial \vec{p}}-
q_iB\tau \left(v_y\frac{\partial f}{\partial p_x}-
v_x \frac{\partial f}{\partial p_y}\right)-
\tau \vec{v} .\frac{\partial f}{\partial \vec{r}}=0.
\label{ansatz_rbte}
\end{eqnarray}
We further calculate $\frac{\partial f}{\partial p_y}$ and
$\frac{\partial f}{\partial p_x}$ using the 
ansatz~\eqref{ansatz} as
\begin{eqnarray}
\left(v_x\frac{\partial f}{\partial p_y}-
v_y \frac{\partial f}{\partial p_x}\right)=
\bigg(v_y \lambda_x + v_y \tau q E_x-
v_x \lambda_y -v_x \tau q E_y\bigg)\frac{\partial f_{T}}
{\partial \omega}\frac{1}{\omega}.
\label{velocity_term}
\end{eqnarray}
Here, we have taken only those terms which are linear in the 
velocity. 
Now substituting Eq.~\eqref{velocity_term} 
into~Eq.~\eqref{ansatz_rbte} and doing some straightforward algebra we arrive at 
\begin{eqnarray}
v_x\left[\frac{\lambda_x}{\tau}-\omega_c \tau q E_y-
\omega_c \lambda_y +\left(\frac{\omega -\mu}{T}\right)\frac{\partial T}
{\partial x}\right]+v_y\left[\frac{\lambda_y}{\tau}+\omega_c \tau q E_x+
\omega_c \lambda_x +\left(\frac{\omega -\mu}{T}\right)\frac{\partial T}
{\partial y}\right]
=0,
\end{eqnarray}
where $\omega_c=\frac{|q_fB|}{\omega}$ corresponds to the cyclotron
frequency. Now, let us equate the coefficient of $v_x$ and $v_y$ 
from both side of the above equation, we get 
\begin{eqnarray}\label{lambda1}
\frac{\lambda_x}{\tau}-\omega_c \tau q_fE_y-
\omega_c \lambda_y +\left(\frac{\omega -\mu}{T}\right)
\frac{\partial T}
{\partial x}=0,\\
\frac{\lambda_y}{\tau}+\omega_c \tau q_fE_x+
\omega_c \lambda_x +\left(\frac{\omega -\mu}{T}\right)
\frac{\partial T}
{\partial y}=0.
\label{lambda2}
\end{eqnarray}
In order to obtain the values of $\lambda_x$ and $\lambda_y$, we solve above Eqs.~\eqref{lambda1} and~\eqref{lambda2} and get
\begin{eqnarray}
\lambda_x =- \frac{\omega_c^2 \tau^3 }
{1+\omega_c^2 \tau^2}q_f E_x -
\frac{\tau}{1+\omega_c^2 \tau^2}\left(\frac{\omega -\mu}{T}\right)
\frac{\partial T}{\partial x}+
\frac{\omega_c \tau^2 }{1+\omega_c^2 \tau^2}q E_y-
\frac{\omega_c \tau^2 }{1+\omega_c^2 \tau^2}\left(\frac{\omega -\mu}{T}\right)
\frac{\partial T}{\partial y},\\
\lambda_y =- \frac{\omega_c \tau^2 }
{1+\omega_c^2 \tau^2}q_f E_x+
\frac{\omega_c \tau^2}{1+\omega_c^2 \tau^2}\left(\frac{\omega -\mu}{T}\right)
\frac{\partial T}{\partial x}-
\frac{\omega_c^2 \tau^3}{1+\omega_c^2 \tau^2}q E_y-
\frac{ \tau}{1+\omega_c^2 \tau^2}\left(\frac{\omega -\mu}{T}\right)
\frac{\partial T}{\partial y}.
\end{eqnarray}
We have exploited $\lambda_x$ and $\lambda_y$ in Eqs.~\eqref{ansatz} to obtain the deviation  $\delta f$ is the quark distribution function as
\begin{eqnarray}
\delta f&=&\{1+(q-1)\beta(\omega-\mu)\}^{-1} \frac{\partial f_{T}}{\partial \omega}\left[-\frac{\tau}
{1+\omega_c^2 \tau^2}q_f v_x+
\frac{\omega_c \tau^2 }{1+\omega_c^2 \tau^2}q_f v_y\right]E_x\nonumber\\
&&+\{1+(q-1)\beta(\omega-\mu)\}^{-1} \frac{\partial f_T}{\partial \omega}\left[-\frac{\tau}
{1+\omega_c^2 \tau^2}q_fv_y
 -\frac{\omega_c \tau^2 }{1+\omega_c^2 \tau^2}q_f v_x\right]E_y\nonumber\\
&&+\{1+(q-1)\beta(\omega-\mu)\}^{-1} \frac{\partial f_T}{\partial \omega}\left[\frac{ \tau}
{1+\omega_c^2 \tau^2}\left(\frac{\omega -\mu}{T}\right) v_x -
\frac{\omega_c \tau^2 }{1+\omega_c^2 \tau^2}
\left(\frac{\omega -\mu}{T}\right)v_y\right]\frac{\partial T}{\partial x}\nonumber\\
&& +\{1+(q-1)\beta(\omega-\mu)\}^{-1} \frac{\partial f_{T}}{\partial \omega}\left[\frac{ \tau}
{1+\omega_c^2 \tau^2}\left(\frac{\omega -\mu}{T}\right) v_y +
\frac{\omega_c \tau^2 }{1+\omega_c^2 \tau^2}
\left(\frac{\omega-\mu}{T}\right)v_x\right]\frac{\partial T}{\partial y}.
\label{delta_f1}
\end{eqnarray}
Following the similar steps, we solve the RBTE for the anti-quarks and get the deviation in the anti-quark distribution function $\delta \bar{f}$ as (replacing $q_i$ with -$q_i$ and $\omega_c$ with 
 - $\omega_c$ in Eq.~\eqref{delta_f1})
\begin{eqnarray}
\delta \bar{f}&=&\{1+(q-1)\beta(\omega +\mu)\}^{-1} \frac{\partial 
\bar{f}_{T}}{\partial \omega}\left[\frac{\tau}
{1+\omega_c^2 \tau^2}q_f v_x+
\frac{\omega_c \tau^2 }{1+\omega_c^2 
\tau^2}q_f v_y\right]E_x\nonumber\\
&&+\{1+(q-1)\beta(\omega +\mu)\}^{-1} \frac{\partial \bar{f}_{T}}{\partial 
\omega}\left[\frac{\tau}
{1+\omega_c^2 \tau^2}q_f v_y
 -\frac{\omega_c \tau^2 }{1+\omega_c^2 
 \tau^2}q_f v_x\right]E_y\nonumber\\
&&+\{1+(q-1)\beta(\omega +\mu)\}^{-1} \frac{\partial \bar{f}_{T}}{\partial 
\omega}\left[\frac{ \tau}
{1+\omega_c^2 \tau^2}\left(\frac{\omega -\mu}{T}\right) v_x +
\frac{\omega_c \tau^2 }{1+\omega_c^2 \tau^2}
\left(\frac{\omega -\mu}{T}\right)v_y\right]
\frac{\partial T}{\partial x}\nonumber\\
&& +\{1+(q-1)\beta(\omega +\mu)\}^{-1} \frac{\partial \bar{f}_{T}}{\partial \omega}
\left[\frac{ \tau}
{1+\omega_c^2 \tau^2}\left(\frac{\omega-\mu}{T}
\right) v_y -
\frac{\omega_c \tau^2 }{1+\omega_c^2 \tau^2}
\left(\frac{\omega -\mu}{T}\right)v_x\right]
\frac{\partial T}{\partial y}.
\end{eqnarray}
 We substitute  $\delta f$ and  $\delta \bar{f}$
obtained here into~\eqref{current_weak} to obtain the $x$ and $y$
components of the induced current density  which are given by  
\begin{eqnarray}\label{J_x}
J_{x}&=&q_fg_f\left[(q_f\beta I_{1})E_x+
(q_f\beta I_{2})E_y
+(\beta^2 I_{3})\frac{\partial T}{\partial x}+
(\beta^2 I_{4})\frac{\partial T}{\partial y}\right],\\
J_{y}&=&q_fg_f\left[(-q_f\beta I_{2})E_x
+(q_f\beta I_{1})E_y
+(-\beta^2 I_{4})\frac{\partial T}{\partial x}+
(\beta^2 I_{3})\frac{\partial T}{\partial y}
\right].
\label{J_y}
\end{eqnarray}
Here the integrals 
$I_1$, $I_2$, $I_3$ and $I_4$ stand for
\ba
I_1=I_{1q}+I_{1\bar{q}}, \\
I_2=I_{2q}+I_{2\bar{q}},\\
I_3=I_{3q}+I_{3\bar{q}},\\
I_4=I_{4q}+I_{4\bar{q}},
\ea
where
\begin{eqnarray*}
I_{1q}&=& \frac{1}{3}\int \frac{d^3p}{(2\pi)^3}~\frac{p^2}{\omega^2}
\frac{ \tau}{(1+\omega_c^2 \tau^2)}~\{1+(q-1)\beta(\omega -\mu)\}^{-1} 
f_T(1-f_T),\\
I_{1\bar{q}}&=&\frac{1}{3}\int \frac{d^3p}{(2\pi)^3}~\frac{p^2}{\omega^2}
\frac{ \tau}{(1+\omega_c^2 \tau^2)}~\{1+(q-1)\beta(\omega +\mu)\}^{-1} 
\bar{f}_T(1-\bar{f}_T),\\
I_{2q}&=&\frac{1}{3}\int \frac{d^3p}{(2\pi)^3}~\frac{p^2}{\omega^2}
\frac{\omega_c \tau^2}{(1+\omega_c^2 \tau^2)}~\{1+(q-1)\beta(\omega -\mu)\}^{-1} 
f_T(1-f_T),\\
I_{2\bar{q}}&=&-\frac{1}{3}\int \frac{d^3p}{(2\pi)^3}~\frac{p^2}{\omega^2}
\frac{\omega_c \tau^2}{(1+\omega_c^2 \tau^2)}~\{1+(q-1)\beta(\omega +\mu)\}^{-1} 
\bar{f}_T(1-\bar{f}_T),\\
I_{3q}&=&-\frac{1}{3}\int \frac{d^3p}{(2\pi)^3}~\frac{p^2}{\omega^2}
\frac{\tau}{(1+\omega_c^2 \tau^2)}~\{1+(q-1)\beta(\omega -\mu)\}^{-1} 
(\omega -\mu)f_T(1-f_T),\\
I_{3\bar{q}}&=&\frac{1}{3}\int \frac{d^3p}{(2\pi)^3}~\frac{p^2}{\omega^2}
\frac{\tau}{(1+\omega_c^2 \tau^2)}~\{1+(q-1)\beta(\omega +\mu)\}^{-1} 
(\omega +\mu)\bar{f}_T(1-\bar{f}_T),\\
I_{4q}&=&-\frac{1}{3}\int \frac{d^3p}{(2\pi)^3}~\frac{p^2}{\omega^2}
\frac{\omega_c \tau^2}{(1+\omega_c^2 \tau^2)}(\omega -\mu)
~\{1+(q-1)\beta(\omega +\mu)\}^{-1} 
f_T(1-f_T),\\
I_{4\bar{q}}&=&-\frac{1}{3}\int \frac{d^3p}{(2\pi)^3}~\frac{p^2}{\omega^2}
\frac{\omega_c \tau^2}{(1+\omega_c^2 \tau^2)}(\omega +\mu)
~\{1+(q-1)\beta(\omega +\mu)\}^{-1} \bar{f}_T(1-\bar{f}_T).
\end{eqnarray*}
The components of the induced current density along the x and y directions disappear at the state of equilibrium. {\em i.e.} $J_{x}=J_{y}=0$. We can write 
from Eqs.~\eqref{J_x} and~\eqref{J_y}
\ba \label{J_x0} 
C_1 E_x+C_2E_y+C_3\frac{\partial T}{\partial x}
+C_4\frac{\partial T}{\partial y}&=&0, \\
-C_2 E_x+C_1E_y-C_4\frac{\partial T}{\partial x}
+C_3\frac{\partial T}{\partial y}&=&0, 
\label{J_y0}
\ea
where $C_1=qI_1$, $C_2=qI_2$,
 $C_3=\beta I_3$, $C_4=\beta I_4$. The matrix equation which links the components of the electric field and temperature gradients to thermoelectric  coefficients, is given by
\ba
  \begin{pmatrix}
  E_x \\
  E_y
  \end{pmatrix}
  =\begin{pmatrix}
  S^B & N|B|\\
  -N|B| & S^B
  \end{pmatrix}
  \begin{pmatrix}
  \frac{\partial T}{\partial x}~\\
  \frac{\partial T}{\partial y}
  \end{pmatrix}.
\ea   
We solve Eqs.~\eqref{J_x0}  and \eqref{J_y0} for 
$E_x$ and $E_y$  
 as 
\ba
E_x=\left(-\frac{C_1C_3+C_2C_4}{C_1^2+C_2^2}\right)
\frac{\partial T}{\partial x} 
+\left(-\frac{C_2C_3-C_1C_4}{C_1^2+C_2^2}\right)
\frac{\partial T}{\partial y},\\
E_y=\left(-\frac{C_1C_3+C_2C_4}{C_1^2+C_2^2}\right)
\frac{\partial T}{\partial y} 
-\left(-\frac{C_2C_3-C_1C_4}{C_1^2+C_2^2}\right)
\frac{\partial T}{\partial x},
\ea
which give the Seebeck and Nernst coefficients 
\ba
S^B &=& -\frac{(C_1C_3+C_2C_4)}{C_1^2+C_2^2},\\
N|B|&=&\frac{(C_2C_3-C_1C_4)}{C_1^2+C_2^2},
\ea
respectively. The integrals $C_2$ and $C_4$ vanishes in the 
absence of the magnetic field, hence Nernst 
coefficient would also disappear.\par
We will now calculate the medium's Seebeck and Nernst coefficients. The x and y components of the current in the medium  can be expressed as the sum of the individual contributions as
\begin{eqnarray}
J_x&=&\sum_{i}\left[q_i (I_1)_iE_x+q_i (I_2)_iE_y
+\beta (I_3)_i\frac{\partial T}{\partial x}+
\beta (I_4)_i\frac{\partial T}{\partial y}\right],\\
J_y&=&\sum_{i}\left[q_i (I_2)_iE_x+q_i (I_1)_iE_y-
\beta (I_4)_i\frac{\partial T}{\partial x}+
\beta (I_3)_i\frac{\partial T}{\partial y}
\right].
\end{eqnarray}
 The 
Seebeck and Nernst coefficients of the medium can be extracted 
by imposing the 
steady state condition ({\em i.e.} putting $J_x=J_y=0$) as
\ba
S^{B}_{tot} &=& -\frac{(K_1K_3+K_2K_4)}{K_1^2+K_2^2},\\
N|B|&=&\frac{(K_2K_3-K_1K_4)}{K_1^2+K_2^2}.
\ea
where
\ba
K_1&=&\sum_{i}q_i (I_1)_i, \quad \quad 
 K_2=\sum_{i}q_i (I_2)_i,\\
K_3&=&\sum_{i}\beta(I_3)_i, \quad \quad  
K_4=\sum_{i}\beta (I_4)_i.
\ea
These expressions for the Seebeck and Nernst coefficients reduces to those obtained for the extensive QCD medium in~\cite{Dey:PRD104'2021}, when the nonextensive 
parameter $q$ approaches to unity. 
\section{Numerical results and discussion}\label{three}
  { 

In this work, we have used a quasi-particle model developed by
Gorenstein and Yang~\cite{Gorenstein:PRD52'1995} to incorporate the strong interaction among the quarks and gluons. The study of transport properties of QCD matter requires an effective description of the strongly interacting medium specially around the QCD transition temperature, where non-perturbative effects play an important role. At asymptotically high temperatures, the thermodynamic and transport properties are studied using the perturbative QCD but the temperature achieved in the heavy ion collision experiments  are only a few times larger than the critical temperature hence use of effective models becomes appropriate. The main idea behind these effective models is that a many particle interacting system like QGP can be described in terms of the non-interacting quasi-particles generated as a result of the collective excitation in the medium. The medium dependent mass of the partons are calculated from the poles of the resummed propagator computed in the Hard Thermal loop (HTL) approximation. In quasi-particle model, the dispersion relation of the quark is given by 
\ba
\omega_i=\sqrt{p^2+m_i^2}
\ea
 where 
 $ 
m_{i}=m_{i,0}+m_{i,T}
$. $m_{i,0}$ and $m_{i,T}$ are the current quark mass and effective thermal mass 
respectively. The thermal mass of the quarks and gluons have been computed using the perturbative thermal QCD as~\cite{Braaten:PRD45'1992,Peshier:PRD66'2002}
\begin{eqnarray}
m_{i,T}^2&=&\frac{g^2T^2}{6}\left(1+\frac{\mu_i^2}{\pi^2T^2}\right).
\end{eqnarray}
}
We have exploited the medium dependent mass in the dispersion relation of quarks to study the Seebeck and Nernst coefficients of the hot and dense QCD medium numerically in the upcoming subsections. In addition, the non-extensive parameter $q$ is taken as an input parameter to study its effects on the
thermoelectric properties of the QGP. The value of $q$ has been 
predicted in the range from $1.1$ to $1.2$ from the transverse momentum distribution of the identical charged particles at RHIC and LHC~\cite{cleymans:JPG39'2012}. In Ref~\cite{Zheng:PRD92'2015}, the value of $q$ has been extracted by fitting the spectra of the particles using the non-extensive distribution function and is found between $1.063$ to $1.25$ for mesons and between 1.05 to 1.25 for protons. Other groups have also obtained the $q$ values in the same range. It has been also argued that the value of $q$ should not be far from 1 in order to maintain the thermodynamic consistency~\cite{Bhattacharya:PRD94'2016,Parvan:EPJA102'2017}. Therefore, we have
studied the Seebeck and Nernst coefficients of the QGP with the values of $q$ in the range $1.1$ to $1.2$. In this study, we have taken $N_c=3$, $N_f=3$ and the relaxation time ($\tau$) for the quarks has been taken from~\cite{Hosoya:NPB250'1985}, which is given by
\ba
\tau(T) =\frac{1}{5.1T \alpha_s^2 \log \left(\frac{1}{\alpha_s}\right)
\label{tau_B0} 
[1+0.12(2N_f+1)]},  
\ea 
where $\alpha_s$ is the running coupling 
constant which is given by
\begin{eqnarray}
\alpha_s (T)=\frac{g'^{2}}{4 \pi}=
\frac{6\pi}{(33-2N_f) \ln \left(\frac{Q}
{\Lambda_{QCD}}\right)},
\label{coupling_T}
\end{eqnarray}
and $Q$ is set at $2 \pi\sqrt{T^2+\frac{\mu^2}{\pi^2}}$.  We have used the similar relaxation time~\eqref{tau_B0} in the presence of magnetic field also since  magnetic field is not a dominant energy scale and it's strength is weak. The magnetic field dependence in the relaxation time enters through the strong coupling constant~\cite{Ayala}
\ba
\alpha_s(\Lambda^2, |q_fB|) = \frac{ \alpha_s(\Lambda^2)  }{1 + b_1\alpha_s(\Lambda^2)\ln\big( \frac{\Lambda^2}{\Lambda^2 + |q_fB|} \big)    },~~\text{where~~} \alpha_s(\Lambda^2) = \frac{1}{ b_1\ln\big( \Lambda^2 / \Lambda^2_{ \overline{\text{MS}} } \big)   },
\label{res-1}
\ea    
here $b_1 = (11N_c - 2N_f)/12\pi$, $\Lambda_{ \overline{ \text{MS} } } = 0.176$ GeV. Now in the upcoming subsections $A$ and $B$, we will discuss the numerical results in absence and in the presence of the magnetic field, respectively.
\newpage
\subsection{In the absence of magnetic field}\label{three1}

Figure~\ref{seebeck_B0}(a) depicts the variation of the Seebeck coefficient with temperature at a fixed chemical potential of $\mu=100$ MeV for different values of the nonextensive parameter $q$. In numerical analysis, we have considered $q=1.1,1.15$ and $1.2$. We have noticed that the Seebeck coefficient decreases monotonically with temperature for all values of $q$, which indicates that the thermoelectric response of the hot QCD matter  becomes progressively weaker in the high-temperature deconfined phase. This behaviour of Seebeck coefficient with temperature can be attributed to enhanced thermal agitation of the charge carriers. At higher temperatures, the particle momentum distribution becomes broader, and the random thermal motion dominates over the directed diffusion induced by the temperature gradient. This leads to a smaller value of Seebeck coefficient.
 Furthermore, increasing the nonextensive parameter from $q=1.0$ to $q=1.2$ consistently lowers the Seebeck coefficient. This behavior originates from the broader power-law momentum distribution associated with Tsallis statistics, which enhances high-energy quark populations and thermal transport while reducing the efficiency of thermoelectric charge separation. The impact  of nonextensivity is visible near the transition temperature and becomes less significant at higher temperatures, where all curves gradually converge. This shows that non-extensivity lowers the thermoelectric response of the hot QCD medium.\\

\begin{center}
\begin{figure}[H]
\begin{tabular}{cc}
\includegraphics[width=7.0cm]{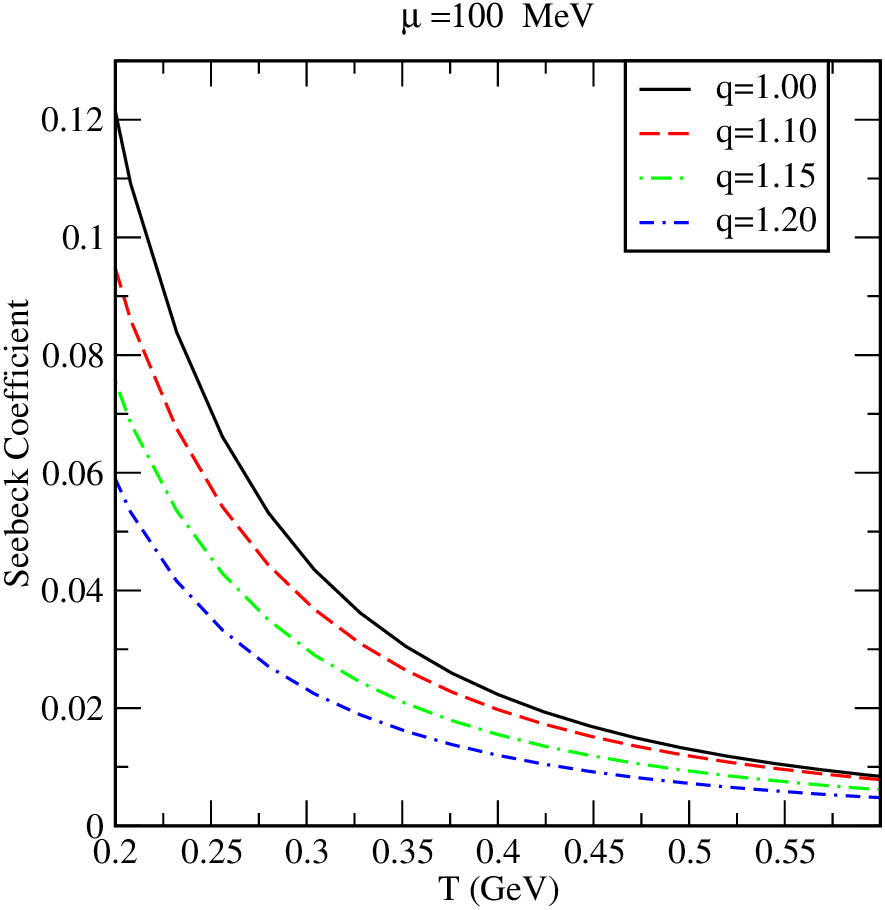}&
\hspace{2.5cm}
\includegraphics[width=7.0cm]{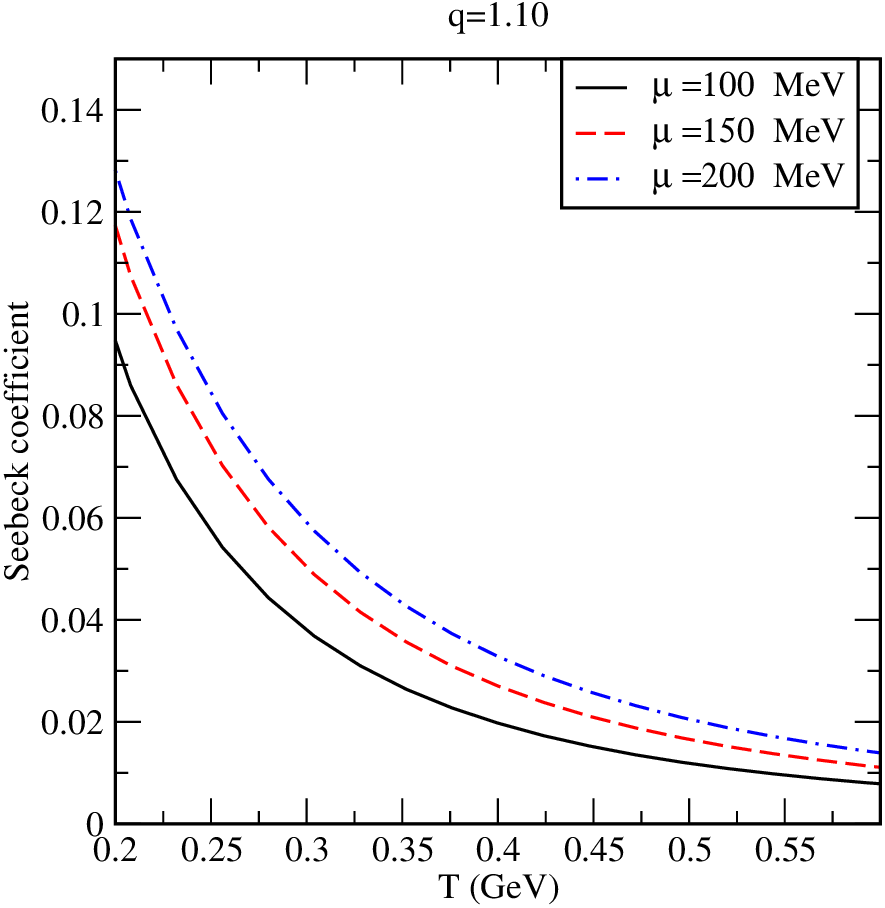}\\
(a)& \hspace{2.5cm} (b) 
\end{tabular}
\caption{{\bf (a)} Seebeck coefficient as a function of temperature at different values of the non-extensive parameter ($q=1.0,1.1,1.15$ and $1.2$). The strength of the chemical potential has been taken as  $\mu=100$ MeV {\bf (b)} Seebeck coefficient as a function of temperature at different values of quark chemical potential ($\mu=100,150 $ and $200$ MeV). The value of non-extensive parameter $q$ is fixed at 1.1.  }
\label{seebeck_B0} 
\end{figure}
\end{center}

Figure~\ref{seebeck_B0}(b) illustrates the temperature dependence of the Seebeck coefficient for three different values of the quark chemical potential, $\mu=100$, $150$, and $200$ MeV at a fixed nonextensive parameter $q=1.1$. We have noticed that the Seebeck coefficients gets enhanced with the quark chemical potential in the temperature range $T=200$ MeV to $T=600$ MeV. 
At large chemical potentials, the density of quark increases while that of anti-quark decreases. This produces an asymmetry between the quark and antiquark number densities which leads to enhanced diffusion of quarks in comparison to the anti-quarks due to the thermal gradient, hence 
a stronger thermoelectric response. 
It is also observed that the effect of the chemical potential is more pronounced in the low temperature region near the transition temperature, while the curves  approach each other in the high temperature region. This indicates that the thermoelectric properties are strongly influenced by the finite baryon density near the crossover region, whereas at sufficiently high temperatures the transport dynamics become predominantly governed by thermal excitations. Consequently, the sensitivity of the Seebeck coefficient to the chemical potential diminishes in the high-temperature deconfined phase.
 The similar trend with temperature and quark chemical potential has been  observed in other studies done for equilibrium QGP described by the Fermi-Dirac distribution~\cite{Dey:PRD102'2020,Dey:PRD104'2021,Khan:PRC110'2024,
Shaikh:PRD11'2025}.
\subsection{In the presence of  magnetic field}\label{three2}
Figure~\ref{seebeck_B3}(a) shows the variation of the Seebeck coefficient with temperature for different values of the nonextensive parameter $q$ in the presence of a weak magnetic field, $eB=0.3\,m_{\pi}^{2}$, at a fixed quark chemical potential $\mu=100$ MeV. The Seebeck coefficient is found to be decreasing with temperature in the presence of the magnetic field also.
\begin{center}
\begin{figure}[H]
\begin{tabular}{cc}
\includegraphics[width=7.0cm]{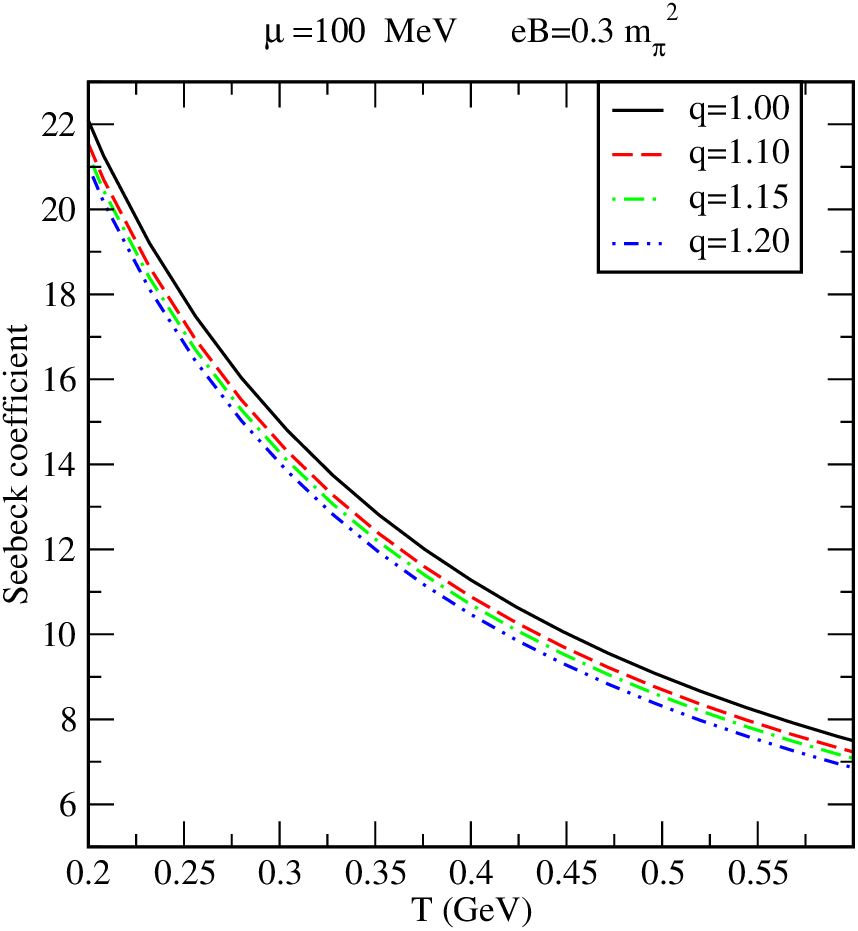}&
\hspace{2.5cm}
\includegraphics[width=7.0cm]{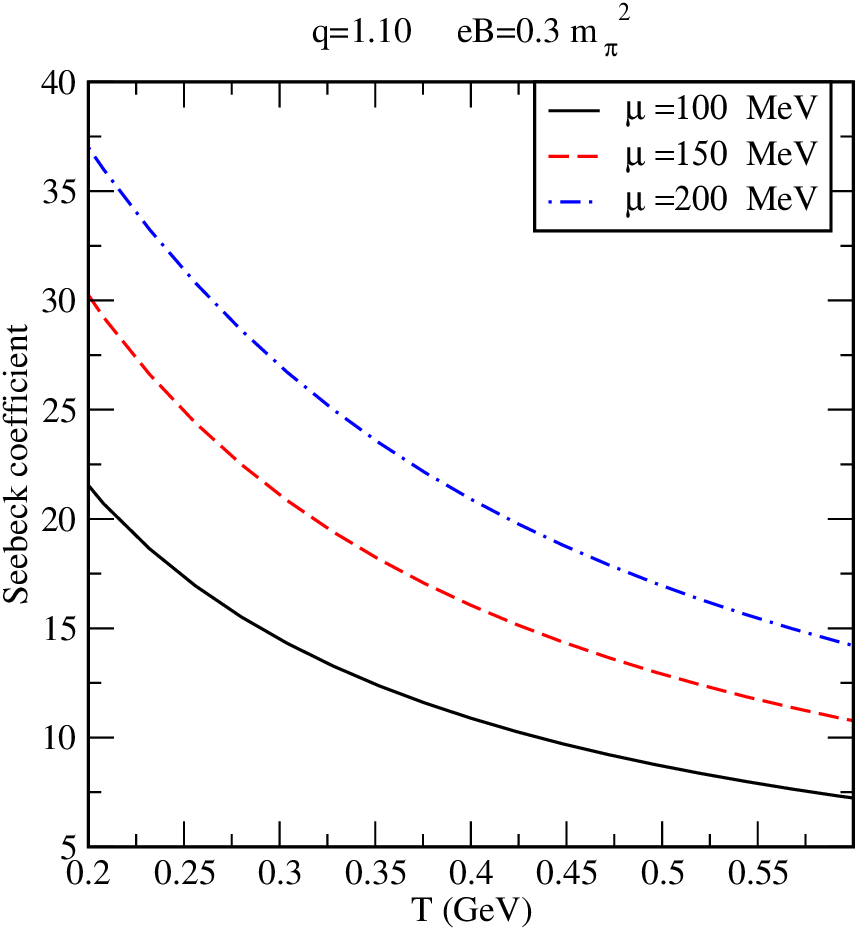}\\
(a)& \hspace{2.5cm} (b) 
\end{tabular}
\caption{{\bf (a)} Seebeck coefficient as a function of temperature at different values of the non-extensive parameter ($q=1.0, 1.1, 1.15$ and $1.20 $ ). The strength of the quark chemical potential and magnetic field is taken as $\mu=100$ MeV and $eB=0.3 m_{\pi}^2$, respectively. {\bf (b)} Seebeck coefficient as a function of temperature at different values of quark chemical potential ($\mu=100,150$ and $200$ MeV). The values of $q$ and $eB$ are fixed at $1.1$ and $0.3 m_{\pi}^2$, respectively. }
\label{seebeck_B3} 
\end{figure}
\end{center}
The effect of nonextensivity remains qualitatively similar in the presence of the magnetic field. It is observed that the Seebeck coefficient decreases systematically with increasing values of the nonextensive parameter,
$
S(q=1.00)>S(q=1.10)>S(q=1.15)>S(q=1.20),
$
throughout the considered temperature range. The reduction originates from the broader momentum distribution associated with Tsallis statistics. As $q$ increases, the high-momentum tail of the distribution becomes increasingly populated, enhancing thermal transport while simultaneously reducing the efficiency of thermoelectric charge separation. 
The presence of the weak magnetic field modifies the motion of charged quarks through the Lorentz force, leading to cyclotron motion in the plane perpendicular to the magnetic field. As a consequence, the transverse diffusion of charge carriers is partially suppressed, which alters the transport coefficients entering the Seebeck coefficient. Nevertheless, since the magnetic field considered here is relatively weak ($eB=0.3\,m_{\pi}^{2}$), its influence is perturbative and does not modify the overall temperature dependence of the Seebeck coefficient. Instead, the dominant behavior continues to be governed by thermal excitations and nonextensive effects.
It is also evident that the separation between different $q$ curves is more pronounced in the low-temperature region, whereas the curves gradually approach one another at higher temperatures. 
This indicates that nonextensive effects are more significant near the QCD crossover region, while at sufficiently high temperatures thermal excitations dominate the transport dynamics, rendering the Seebeck coefficient progressively less sensitive to deviations from equilibrium.\par

In figure~\ref{seebeck_B3}(b), we have shown  the temperature dependence of the Seebeck coefficient for three different quark chemical potentials, $\mu=100$, $150$, and $200$ MeV, in the presence of a weak magnetic field, $eB=0.3\,m_{\pi}^{2}$, at a fixed nonextensive parameter $q=1.1$. We have observed the same increasing pattern of the Seebeck coefficient with quark chemical potential  as in the absence of the magnetic field.
The figure further shows that the Seebeck coefficient increases significantly with increasing chemical potential throughout the considered temperature range,
$
S(\mu=200~\mathrm{MeV}) >
S(\mu=150~\mathrm{MeV}) >
S(\mu=100~\mathrm{MeV}).
$
 Nevertheless, because the magnetic field strength considered here is relatively small ($eB=0.3\,m_{\pi}^{2}$), its effect is insufficient to alter the overall temperature dependence of the Seebeck coefficient. Instead, the dominant trends continue to be governed by thermal excitations and the finite baryon density of the medium.
\par

Figure~\ref{seebeck_B4}(a) depicts the variation of the Nernst coefficient with temperature for different values of the nonextensive parameter $q$ in the presence of a weak magnetic field at a fixed quark chemical potential. The Nernst coefficient exhibits a pronounced temperature dependence, indicating that the transverse thermoelectric response of the quark-gluon plasma (QGP) is highly sensitive to both thermal effects and deviations from equilibrium.
The Nernst effect originates from the combined action of a temperature gradient and an external magnetic field. In the presence of a magnetic field, the Lorentz force deflects the thermally diffusing charged quarks, resulting in the generation of a transverse electric field. Therefore, unlike the Seebeck effect, which is purely longitudinal, the Nernst coefficient is governed by the interplay between thermal diffusion and the Hall motion of the charge carriers.

\begin{center}
\begin{figure}[H]
\begin{tabular}{cc}
\includegraphics[width=7.0cm]{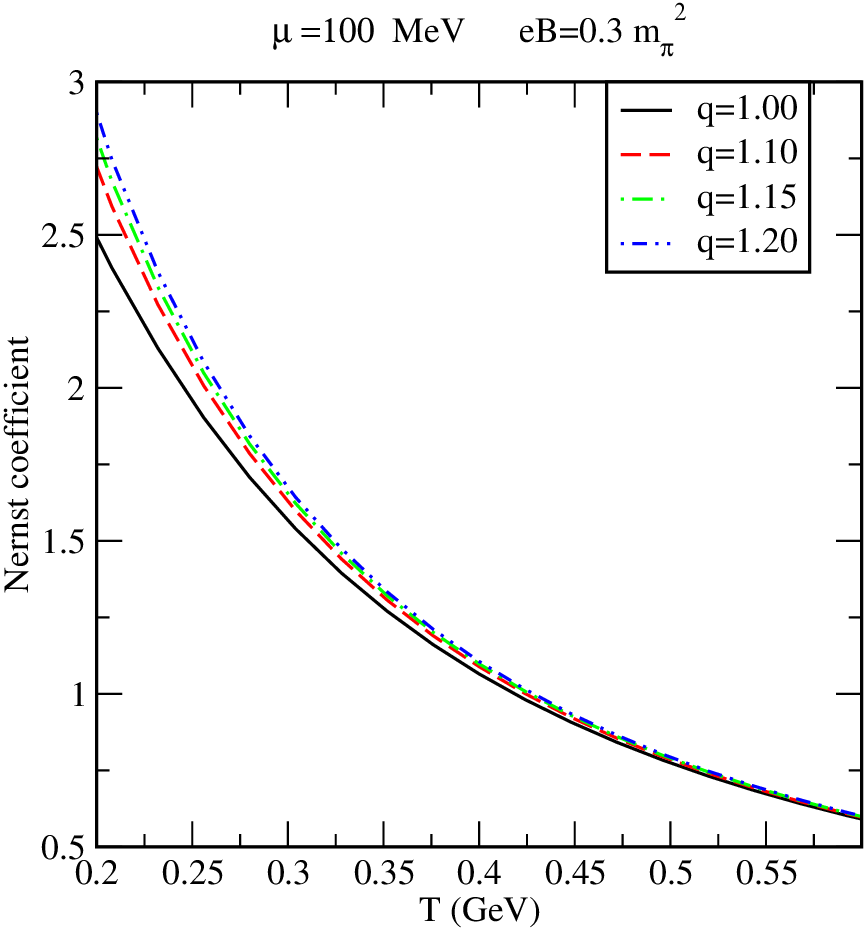}&
\hspace{2.5cm}
\includegraphics[width=7.0cm]{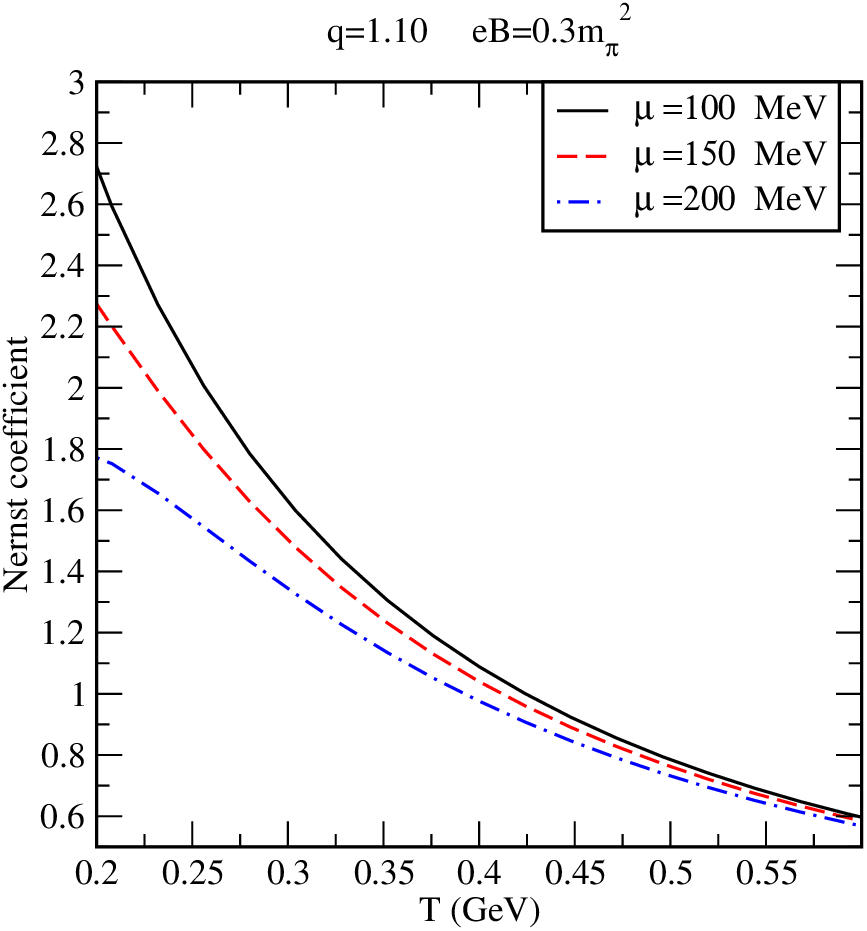}\\
(a)& \hspace{2.5cm} (b) 
\end{tabular}
\caption{{\bf (a)} Nernst coefficient as a function of temperature at different values of the non-extensive parameter ($q=1.0, 1.1, 1.15$ and $1.20 $ ). The strength of the quark chemical potential and magnetic field is taken as $\mu=100$ MeV and $eB=0.3 m_{\pi}^2$, respectively. {\bf (b)} Nernst coefficient as a function of temperature at different values of quark chemical potential ($\mu=100,150$ and $200$ MeV). The value of $q$ and $eB$ is fixed at $1.1$ and $0.3 m_{\pi}^2$, respectively. }
\label{seebeck_B4} 
\end{figure}
\end{center}

As the temperature increases, the magnitude of the Nernst coefficient decreases monotonically. At relatively low temperatures, the charge carriers possess longer relaxation times and experience a more effective deflection by the magnetic field, giving rise to a stronger transverse thermoelectric response. However, with increasing temperature, frequent partonic scatterings shorten the relaxation time and randomize the motion of quarks. Consequently, the Hall deflection becomes less effective, suppressing the transverse electric field generated by the temperature gradient. This leads to a gradual reduction of the Nernst coefficient at higher temperatures.

The figure~\ref{seebeck_B4}(a) further demonstrates that the Nernst coefficient increases  systematically with increasing values of the nonextensive parameter,
$
N|B|(q=1.20)>
N|B|(q=1.15)>
N|B|(q=1.10)>
N|B|(q=1.00),
$
throughout the temperature range. 
It is also observed that the separation between different $q$ curves is more significant in the low-temperature region and gradually diminishes with increasing temperature. This behavior indicates that nonextensive effects play a more prominent role near the QCD crossover region, whereas thermal excitations dominate the transport dynamics in the high-temperature deconfined phase. Consequently, the sensitivity of the Nernst coefficient to deviations from equilibrium becomes weaker at sufficiently high temperatures.
Overall, these results indicate that the transverse thermoelectric response of the magnetized QGP is strongest in the low-temperature regime and is progressively suppressed by both increasing temperature and increasing nonextensivity. The observed behavior reflects the combined influence of thermal fluctuations, magnetic deflection, and non-equilibrium effects on charge transport in the deconfined medium.\par
Figure~\ref{seebeck_B4}(b) illustrates the temperature dependence of the Nernst coefficient for three different values of the quark chemical potential, $\mu=100$, $150$, and $200$ MeV, at a fixed nonextensive parameter $q=1.10$ in the presence of a weak magnetic field, $eB=0.3\,m_{\pi}^{2}$. It is observed that the Nernst coefficient decreases monotonically with increasing temperature for all values of the chemical potential. Furthermore, for a fixed temperature, the magnitude of the Nernst coefficient decreases systematically with increasing chemical potential,
$
N|B|(\mu=100~\mathrm{MeV})
>
N|B|(\mu=150~\mathrm{MeV})
>
N|B|(\mu=200~\mathrm{MeV}).
$
%

\begin{center}
\begin{figure}[H]
\begin{tabular}{cc}
\includegraphics[width=7.0cm]{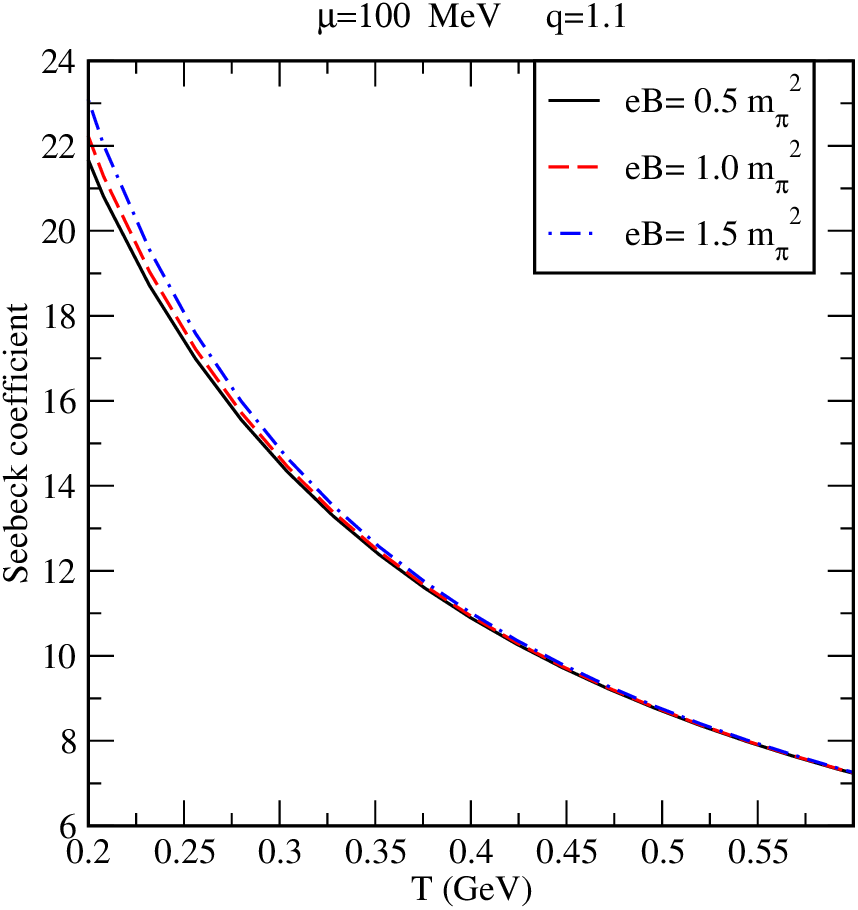}&
\hspace{2.5cm}
\includegraphics[width=7.0cm]{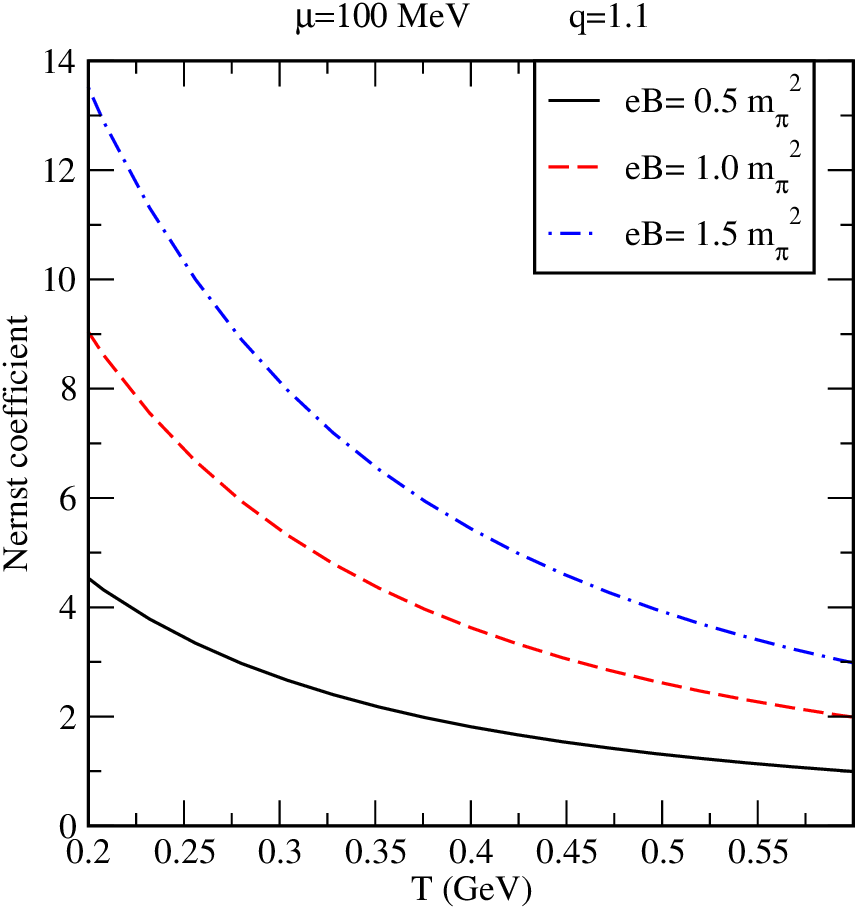}\\
(a)& \hspace{2.5cm} (b) 
\end{tabular}
\caption{{\bf (a)} Seebeck coefficient as a function of temperature at different values of magnetic field ($eB=0.5 m_{\pi}^2,1.0 m_{\pi}^2$ and $1.5 m_{\pi}^2$ ). The values of $\mu$ and $q$ have been taken as 100 MeV and 1.1, respectively. {\bf (b)} Nernst coefficient as a function of temperature at different values of magnetic field ($eB=0.5 m_{\pi}^2,1.0 m_{\pi}^2$ and $1.5 m_{\pi}^2$ ). The values of $\mu$ and $q$ have been taken as 100 MeV and 1.1, respectively. }
\label{seebeck_B5} 
\end{figure}
\end{center}

Figure~\ref{seebeck_B5}(a) illustrates the temperature dependence of the Seebeck coefficient for three different magnetic field strengths at a fixed chemical potential $\mu=100$ MeV and non-extensive parameter $q=1.1$. It is evident that the Seebeck coefficient decreases monotonically with increasing temperature for all values of the magnetic field considered.
At low temperatures ($T\approx0.2$ GeV), the Seebeck coefficient exhibits a noticeable dependence on the magnetic field strength. The magnitude of the Seebeck coefficient increases with increasing magnetic field, satisfying
$
S(eB=1.5\,m_\pi^2)>
S(eB=1.0\,m_\pi^2)>
S(eB=0.5\,m_\pi^2).
$
This enhancement indicates that the magnetic field strengthens the thermoelectric response of the medium in the low-temperature region. As the temperature increases, the difference between the three curves gradually diminishes. Above $T\gtrsim0.45$ GeV, the curves almost overlap, indicating that the influence of the magnetic field becomes increasingly weak at higher temperatures. In this regime, thermal effects dominate over the magnetic-field-induced modifications, making the Seebeck coefficient nearly insensitive to the magnetic field strength.
Overall, the figure demonstrates that while the magnetic field enhances the thermoelectric response at lower temperatures, its effect is progressively suppressed as the quark-gluon plasma approaches the high-temperature regime.
\par

In figure~\ref{seebeck_B5}(b), we have plotted  the Nernst coefficient as a function of temperature for three different magnetic field strengths at a fixed chemical potential $\mu=100$ MeV and non-extensive parameter $q=1.1$. We notice that the Nernst coefficient decreases monotonically with increasing temperature for all values of the magnetic field considered. The decrease is relatively steep in the low-temperature region, while it becomes more gradual at higher temperatures.
Unlike the Seebeck coefficient, the Nernst coefficient exhibits a pronounced dependence on the magnetic field throughout the entire temperature range. For a given temperature, the Nernst coefficient increases significantly with increasing magnetic field strength, following the ordering
$
N|B|(eB=1.5\,m_\pi^2)>
N|B|(eB=1.0\,m_\pi^2)>
N|B|(eB=0.5\,m_\pi^2).
$
This behavior indicates that the transverse thermoelectric response is strongly enhanced in the presence of an external magnetic field.

Although the Nernst coefficient decreases with temperature for all magnetic field strengths, the separation between the curves remains appreciable even at higher temperatures. This suggests that the magnetic field continues to play an important role in governing the transverse transport properties of the quark-gluon plasma, in contrast to the Seebeck coefficient where the magnetic field dependence becomes weaker at high temperatures.
The enhancement of the Nernst coefficient with increasing magnetic field originates from the magnetic-field-induced corrections to the transport coefficients incorporated within the weak-field kinetic theory. As the temperature increases, thermal effects reduce the magnitude of the Nernst coefficient; however, the magnetic field continues to exert a significant influence on the transverse thermoelectric transport, leading to a noticeable separation between the curves even at high temperatures.

\section{Summary and future outlook}\label{four}
{ 

In this work, we have computed the Seebeck and Nernst coefficients of a weakly magnetized nonextensive quark–gluon plasma within the framework of relativistic kinetic theory using the relaxation time approximation. The interactions among the partons  have been incorporated through a quasiparticle model, where the medium dependent quark masses are obtained from the poles of the resummed quark propagator. In order to include the effects of non-extensivity, we have exploited the Tsallis distribution which enables us to explore the influence of non-equilibrium conditions on the thermoelectric transport properties of the hot and dense QCD matter.
Our results shows that the Seebeck coefficient decreases with the non-extensive parameter both in the absence and in the presence of an external magnetic field. This reduction  can be attributed to the fact that  the quark momentum distribution gets modified because of the the non-extensive statistics. On the other hand, the Nernst coefficient, which appears  due to the presence of  magnetic field gets enhanced with increasing non-extensivity. 
} 
The electric current in
the presence of Seebeck effect becomes $J=\sigma_{el}E
-\sigma_{el}S\nabla T$, while thermal conductivity 
gets modified as  $\kappa=
\kappa_0-T\sigma_{el}S^2$. The second law of thermodynamics predicts positive values for the electrical and thermal conductivity. Hence it is important to take into account the Seebeck coefficient of the medium to get more realistic estimates of the electric current and thermal conductivity of the medium. The Seebeck coefficient will also affect the entropy production in the medium so it must be taken into account for calculation of the entropy 
production, which has been 
neglected in~\cite{Gavin:NPA435'1985,Huang:PRD81'2010}. In future, we plan to study the effects of momentum anisotropy
on the various transport coefficients of the non-extensive 
QCD matter. Effect of time and space dependent electromagnetic fields is another project we want to pursue in the future.

\section*{Acknowledgements}
 S. A. K. is thankful to Integral University for providing the necessary facilities for research and assigning the Manuscript Communications No. $IU/R \& D/2026- MCN0004030$.

\appendix
\appendixpage
\addappheadtotoc
\begin{appendix}

\renewcommand{\theequation}{A.\arabic{equation}}
\section{Boltzmann Equation in the weak magnetic field}
The RBTE~\eqref{rbte} can be written in the relaxation time approximation as
\begin{eqnarray}
p^{\mu}\frac{\partial f}{\partial x^{\mu}}+q~F'^{\sigma}
\frac{\partial f}{\partial p^{\sigma}}= 
-p^{\mu}u_{\mu}\nu \left(f- f_{T}
\right)
\label{rbte_weak}
\end{eqnarray}
where $F'^{\sigma}=qF^{\sigma \rho}p_{\rho}=
(p^0 \vec{v}.\vec{F},p^0\vec{F})$, is the covariant form
 of the Lorenz force 
$\vec{F}=q(\vec{E}+\vec{v}\times \vec{B})$. We can write
Eqn.~\eqref{rbte_weak} using $F^{0i}=-E^{i}$ and 
$2F_{ij}=\epsilon_{ijk}B^k$ ($\epsilon_{ijk}$ is 
 anti-symmetric Levi-Civita tensor) as
\begin{eqnarray}
\frac{\partial f}{\partial t}+\vec{v}.
\frac{\partial f}{\partial \vec{r}}+
\frac{\vec{F}.\vec{p}}{p^0}
\frac{\partial f}{\partial p^0}+\vec{F}.
\frac{\partial f}{\partial \vec{p}}=
-\nu \left(f- f_{T}
\right)
\label{B_1}
\end{eqnarray}
considering $p^0$ as an independent variable
\begin{eqnarray}
\frac{\partial}{\partial \vec{p}} \rightarrow
\frac{\partial p^0}{\partial \vec{p}}\frac{\partial}{\partial p^0}+
\frac{\partial}{\partial \vec{p}}=
\frac{\vec{p}}{p^0}\frac{\partial}{\partial p^0}+
\frac{\partial}{\partial \vec{p}}
\end{eqnarray} 
Eqn.~\eqref{B_1} takes the form
\begin{eqnarray}
\vec{v}.
\frac{\partial f}{\partial \vec{r}}+
\vec{F}.\frac{\partial f}{\partial \vec{p}}=
-\nu \left(f-f_{T}
\right)
\end{eqnarray}

\renewcommand{\theequation}{B.\arabic{equation}}

\end{appendix}

\end{document}